\documentclass[aps,pre,reprint,10pt,twocolumn,showpacs,nofootinbib,superscriptaddress]{revtex4-1}
\usepackage{amsmath}
\usepackage{amsfonts}
\usepackage{color}
\usepackage{amssymb}
\usepackage{graphicx}
\usepackage{epstopdf}
\usepackage{comment}

\newcommand{\lrangle}[1]{\langle {#1} \rangle}

\usepackage[normalem]{ulem}

\begin{document}

\title{Continuous and discontinuous absorbing-state phase transitions on Voronoi-Delaunay random lattices}
\author{Marcelo M. de Oliveira}\thanks{On leave at:
Theoretical Physics Division, School of Physics and Astronomy,
The University of Manchester, Manchester, M13 9PL,
UK}

\address{Departamento de F\'{\i}sica e Matem\'atica, Universidade Federal de S\~ao Jo\~ao Del Rei, 36420-000, Ouro Branco, MG, Brazil}

\author{Sidiney G. Alves}\thanks{Present address: Departamento de F\'{\i}sica e Matem\'atica, Universidade Federal de S\~ao Jo\~ao Del Rei, 36420-000, Ouro Branco, MG, Brazil}
\author{Silvio C. Ferreira}
\address{Departamento de F\'isica, Universidade Federal de Vi\c cosa,
36570-000, Vi\c cosa, MG, Brazil}

\date{\today}

\begin{abstract}
We study absorbing-state phase transitions in two-dimensional
Voronoi-Delaunay (VD) random lattices with quenched coordination
disorder. Quenched randomness usually changes the criticality and
destroys discontinuous transitions in low-dimensional nonequilibrium
systems. We performed extensive simulations of the Ziff-Gulari-Barshad
(ZGB) model, and verified that the VD disorder does not change the
nature of its discontinuous transition. Our results corroborate recent
findings of Barghatti and Vojta [Phys. Rev. Lett. {\bf 113}, 120602
(2014)] stating the irrelevance of topological disorder in a class of
random lattices that includes VD and raise the interesting possibility
that disorder in nonequilibrium APT may, under certain conditions, be
irrelevant for the phase coexistence. We also verify that the VD
disorder is irrelevant for the critical behavior of models belonging
to the directed percolation and Manna universality classes.

\end{abstract}

\pacs{02.50.Ey, 05.70.Ln, 64.60.Ht, 75.40.Mg}

\maketitle

\section{Introduction}

Nonequilbrium phase transitions from an active (fluctuating) to an inactive
(absorbing) phase in spatially extended systems is a topic of broad
interest~\cite{Marrobook,henkel08,odor04}. The so-called absorbing-state
phase transitions (APTs) arise in a wide variety of problems as, for
example, heterogeneous catalysis~\cite{zgb}, interface growth~\cite{tang},
population dynamics and epidemiology~\cite{pastor2014}. Recent experimental
realizations in turbulent liquid crystals~\cite{take07}, driven
suspensions~\cite{pine} and superconducting vortices~\cite{okuma} highlight the
importance of this kind of transition.

In analogy with equilibrium phase transitions, it is expected that {\em
continuous} APTs can be classified in  universality
classes~\cite{Marrobook,henkel08}. Generically, single-component systems
with short-range interactions exhibiting a continuous APT, in the absence
of extra symmetries or conservation laws, belong to the directed
percolation (DP) universality class~\cite{gras,jans}, but other 
robust classes emerge when multiple absorbing states and conservation 
laws are included~~\cite{Marrobook,henkel08}.

Of particular interest is how spatially quenched disorder affects the
critical behavior of an APT.  In real systems, quenched disorder
appears in the form of impurities and defects~\cite{hinri00b}. On a
regular lattice, quenched disorder can be added in the forms of random
deletion of sites or bonds~\cite{noestPRL, noestPRB, adr-dic96,
adr-dic98, vojta06, DeOliveira2008} or of random spatial variation of
the control parameter~\cite{durrett, salinas08, vojta14b}.  In all the
cases above, quenched randomness produces rare regions which are
locally supercritical even when the whole system is subcritical. The
lifetime of active rare regions  is exponentially long in the domain
size. The convolution of rare region and exceedingly large lifetimes
can lead to a slow dynamics, with non universal exponents, for some
interval of the control parameter $\lambda_c^{(0)}<\lambda<\lambda_c$
where $\lambda_c^{(0)}$ and $\lambda_c$ are the critical points of the
clean and disordered systems, respectively. This interval of
singularities is called Griffiths phase (GP)~\cite{vojta06b}. This GP
behavior was verified in DP models with uncorrelated disorder
irrespective to the disorder strength and corresponds to the
universality class of the random transverse Ising
model~\cite{HooyberghsPRL,HooyberghsPRE,vojta05,DeOliveira2008,vojta09}.

These findings are in agreement with the heuristic Harris' criterion
\cite{harris74}, which states that uncorrelated quenched disorder is a
relevant perturbation if
\begin{equation}
d\nu_\perp<2,
\end{equation}
where $d$ is the dimensionality and $\nu_\perp$ is the correlation
length exponent of the clean model. Note that in DP this inequality is
satisfied for all dimensions $d<4$, since $\nu_\perp =$ 1.096854(4),
0.734(4) and 0.581(5), for $d=1$, $2$ and $3$, respectively
\cite{jensen92,jensen99,voigt97}. In the opposite way, simulations of
the continuous APT in models with a conserved field in  the Manna
universality class~\cite{manna}, considering uncorrelated lattice
dilution below the lattice percolation threshold, provide strong
evidences that this kind of disorder is irrelevant although the Harris
criterion is satisfied for $d<4$~\cite{LeePRE2011,LeePRE2013,LeePRL}.

For equilibrium {\em discontinuous} phase transitions the Imry-Ma
criterion \cite{ma, ma2} governs the stability of macroscopic phase
coexistence and disorder destroys phase coexistence by domain
formation in dimensions $d \leq 2$. If the distinct phases are related
by a continuous symmetry the marginal dimension is $d = 4$ \cite{ma2}.
Therefore, first-order phase transitions become rounded in presence of
disorder for $d \leq 2$.

Recent numerical results provide evidences that the Imry-Ma argument
for equilibrium systems can be extended to non-equilibrium APTs:
Irrespective to the uncorrelated disorder strength, Buendia and
Rikvold \cite{buendia,BuendiaPhyA,BuendiaPRE} reported that the
absorbing discontinuous transition in the Ziff-Gulari-Barshad (ZGB)
model for heterogeneous catalysis turns to a continuous one (see also
the discussion in \cite{BustosPRE}). Analogous behavior was observed
more recently by Mart\'in {\it et al.} \cite{martin} for a
two-dimensional quadratic contact-process~\cite{Liu2007}.

Another important question is the role played by disorder inherent to the
underlying connectivity in a nonperiodic, random structure of integer
dimension as the random lattice generated by the Voronoi-Delaunay (VD)
triangulation~\cite{hilh08}. This random lattice can be generated  from a
random (uniform) distribution of $N$ points in a unitary square region.
The triplets that can be circumscribed by a circle that does not overlap
with any other point form a triangulation. The result is a two-dimensional
connected graph with a Poissonian distribution of connectivity with average
degree $\bar{q}=6$~\cite{okabe}. This lattice plays an important role in
the description of idealized statistical geometries such as planar cellular
structures, soap throats, {\it etc}.~\cite{okabe,hilh08}.

Recently, it was found that such a kind of VD disorder does not alter the
character of the APT exhibited by the clean contact process (CP)
\cite{oliveira2}, which is a prototypical model in the DP universality
class. These results are in evident contrast with those for
uncorrelated disorder which leads to an infinite-randomness critical
point and strong GPs~\cite{DeOliveira2008,vojta09}.
In order to determine the relevance of the disorder in these cases, we can
apply the heuristic Harris-Luck criterion~\cite{luck93}, in which
the regular critical behavior remains unchanged when the wandering exponent\footnote{The wandering exponent is associated to the decay of
deviations from the average as a function of patch sizes where the averages are
computed.}
$\omega$ does not exceed a threshold value
given by
\begin{equation}
\omega_c=1-\frac{1}{d\nu_\perp}.
\end{equation}
For independent dilution, $\omega=1/2$, and Luck's expression reduces to
the Harris criterion. 

Former numerical estimates of wandering exponents for VD
triangulations indicated a value close to independent dilution
$\omega=1/2$~\cite{janke}. So, the clean critical behavior observed
for CP on VD lattices posed doubts on the validity of Harris-Luck
criterion for DP class~\cite{oliveira2}. This inconsistency was
recently unfolded~\cite{vojta14c} with the determination of the
correct  wandering exponent of VD lattices as $\omega=1/4$ in $d=2$
implying a criterion $\nu_\perp>2/3$, not $\nu_\perp>1$, for a clean
critical behavior.

In the present work, we investigate the role played by the disorder of
VD lattice on the phase coexistence of ZGB model. We provide evidences
that the VD topological disorder does not destroy the phase
coexistence and thus permit discontinuous phase transitions.
We complement the paper with more evidences for the irrelevance of VD
disorder for continuous APTs belonging to the DP~\cite{henkel08} and
Manna~\cite{Turcotte1999,manna} universality classes.

The reminder of this paper is organized as follows. In the next
section, we review the models definitions and details of the
simulation methods we used. In Sec. III, we present our results and
discussions. Sec. IV is devoted to summarize our conclusions.

\section{Models and methods} 
\label{models}
We constructed the Voronoi-Delaunay lattice with periodic boundary
conditions, following the method described in \cite{Friedberg1984}.
For sake of simplicity, the length of the domain where $N$ node  are
randomly distributed will be expressed in terms of $L=\sqrt{N}$.

\subsection{Discontinuous APT:}

The ZGB model~\cite{zgb}, a lattice gas model introduced to
investigate the reaction of CO oxidation on a catalytic substrate, follows the
Langmuir-Hinshelwood mechanism,
\begin{eqnarray}
\mbox{CO}_{gas}+*\to \mbox{CO}_{ads} \nonumber \\
\mbox{O}_{2 gas}+2* \to 2\mbox{O}_{ads} \nonumber \\
\mbox{CO}_{ads}+\mbox{O}_{ads}\to \mbox{CO}_{2}+ 2*, \nonumber
\end{eqnarray}
where $*$ denotes an empty site, and subscripts indicate the state
(gaseous or adsorbed) of each species. The O$_{2 gas}$ dissociates at
surface, and requires two empty sites to adsorb, while CO requires
only one site to adsorb (the model is also called the monomer-dimer
model). The product CO$_2$ desorbs immediately on formation.
CO$_{gas}$ molecules arrive at rate $Y$ per site while O$_2$ arrives
at rate $(1-Y)$, with $0\leq Y\leq 1$. Varying the control parameter
$Y$, the model exhibits phase transitions between an active steady
state and one of the two absorbing or ``poisoned'' states, in which
the surface is saturated either by oxygen (O) or by CO. The first
transition (O-poisoned) is found to be continuous while the second
(CO-poisoned) is strongly discontinuous.

The computer implementation is the following: With a probability $Y$ a
CO adsorption attempt takes place, and with a complementary
probability ($1-Y$) an O$_2$ adsorption attempt takes place. In the
former case, one site is randomly chosen. If the site is occupied,
either by O or CO, the attempt fails. If it is empty but one of its
first-neighbors is occupied by an O, both sites become empty (O and CO
react instantaneously). Otherwise, the site becomes occupied by an
adsorbed CO molecule. Analogous procedure is followed for an O$_2$
adsorption attempt, but in this case we have to choose at random a
$pair$ of first-neighbors sites, and check for the opposite species in
all remaining nearest-neighbors of the target pair.

\subsection{Continuous APT:}

The CP~\cite{harris-CP} is the prototypical model of the DP class, and
is defined on a lattice with each site either active ($\sigma_i=1$) or
inactive ($\sigma_i=0$). Transitions from active to inactive occur
spontaneously at a rate of unity. The transition from inactive to
active occurs at rate $\sigma_j\lambda/k_j$, for each edge between
active nearest neighbors $j$ of site $i$. The computer implementation
of CP in graphs with arbitrary connectivity is as
follows~\cite{Marrobook,DeOliveira2008}: An occupied site is chosen at
random. With probability $p=1/(1+\lambda)$ the chosen particle is
removed. With the complementary probability $1-p=\lambda/(1+\lambda)$,
a nearest neighbor of the selected particle is randomly chosen and, if
empty, is occupied, otherwise nothing happens and simulations runs to
the next step. Time is incremented by $\delta t = 1/n$, where $n$ is
the number of particles. So, the creation mechanism in  CP effectively
compensates the local connectivity variation with a reduction of the
spreading rate through a particular edge inversely proportional to the
connectivity of the site that transmits a new particle. If we modify
these rules to create offspring in {\em all} empty nearest neighbors of
the randomly chosen occupied site we obtain the A
model~\cite{a1,dic-jaf} (in the A-model occupied sites becomes empty
at unitary rate, as in the CP). This means that sites with higher
coordination number produce more activity when compared with CP,
enhancing possible ``rare region effect'' \cite{vojta06b,vojta14}. 
Since contagion occurs more readily in the A model than in the CP, the
critical creation rate $\lambda_c$ is smaller but the two models share
the same critical behavior of the DP universality class~\cite{a2}.

The Manna model~\cite{manna}, a prototypical model of the Manna
class and introduced to investigate the dynamic of sandpiles in the
context of self-organized criticality, is defined on a lattice where
each site assumes integer values (mimicking the number of ``sand
grains" deposited on the substrate). In the version we investigate an
unlimited number of particles per site is permitted. Sites with a
number bellow a threshold $N_c=2$ are inactive while those where this
number is equal to or larger than $N_c$ are active. The active sites
redistribute their particles among its nearest neighbors chosen at
random, generating a dynamics that conserves the number of particles
when considering periodic boundary condition. The Manna model exhibits
a continuous phase transition from an active to an inactive state
depending on the control parameter $p$ that is given by the density of
particles on the lattice~\cite{Dickman2001}. The absorbing stationary
state, where all sites have a number of particles bellow $N_c$ is
characterized by an infinite number of configurations. The computer
implementation is analogous to that of CP: one active site $i$
($N_i\geqslant N_c$) is randomly chosen. Each of the $N_i$ particles
is sent to a randomly  chosen nearest neighbor irrespective of its
state. The site $i$ becomes empty (inactive) and the nearest neighbors
with $N_j\equiv N_c-1$ particles that received a new one are
activated.

\subsection{Simulation methods}

The central method we used involves the quasi-stationary state, in which
averages are restricted to samples that did not visit an absorbing
state~\cite{Marrobook}. To perform the QS analysis we applied the
simulation method of Ref.~\cite{qssimPRE}. The method is based in
maintaining, and gradually updating, a set of configurations visited
during the evolution; when a transition to the absorbing state is
imminent, the system is instead placed in one of the saved
configurations. Otherwise the evolution is exactly that of a
conventional simulation~\cite{qssimPhysA}. Each realization of the
process is initialized in an active state, and runs for at least
$10^8$ Monte Carlo time steps. Averages are taken in the QS regime,
after discarding an initial transient of $10^7$ time steps or more.
This procedure is repeated for each realization of disorder. The
number of disorder realization ranged from 20 (for the largest size
used, $L=2048$) to $10^3$. Another important quantity is the lifetime
in the QS regime, $\tau$. In QS simulations we take $\tau$ as the mean
time between successive attempts to visit the absorbing state.

For discontinuous APTs, we estimated the transition point 
through the jump in the order parameter and the finite-size
scaling of the maximum of the susceptibility. In DP class the spreading analysis starting from a
single active site (a pre-absorbing configuration) is very accurate
and computationally efficient method~\cite{Marrobook}. For Manna
class, spreading analysis is more cumbersome~\cite{henkel08} due to
infinitely many pre-absorbing configurations. So, we proceeded using
dimensionless moment ratios analysis in the QS
state~\cite{dic-jaf}, which are size-independent at criticality. 
Here, we analyze the critical moment ratio $m = \langle \rho^2
\rangle/\langle \rho \rangle ^2$, which assumes a universal value
$m_c$ at the clean critical point.

\section{Results}

\subsection{Discontinuous APT}

First order transitions are characterized by a discontinuity in the
order parameter and thermodynamic densities, with an associated
delta-peak behavior in the susceptibility~\cite{henkel08}. However, at
finite volume thermodynamic quantities become continuous and rounded.
According to the finite-size theory,  rounding and shifting of  the
coexistence point scale inversely proportional to the system volume
$L^d$~\cite{binder}. Although there is no established similar scaling
theory for nonequilibrium systems yet, some studies show evidences of
an analogous behavior for APTs~\cite{AliSaif2009,Sinha2012,DeOliveira2015}.

\begin{figure}[htb]
\begin{center}
\includegraphics[width=8cm,clip=true]{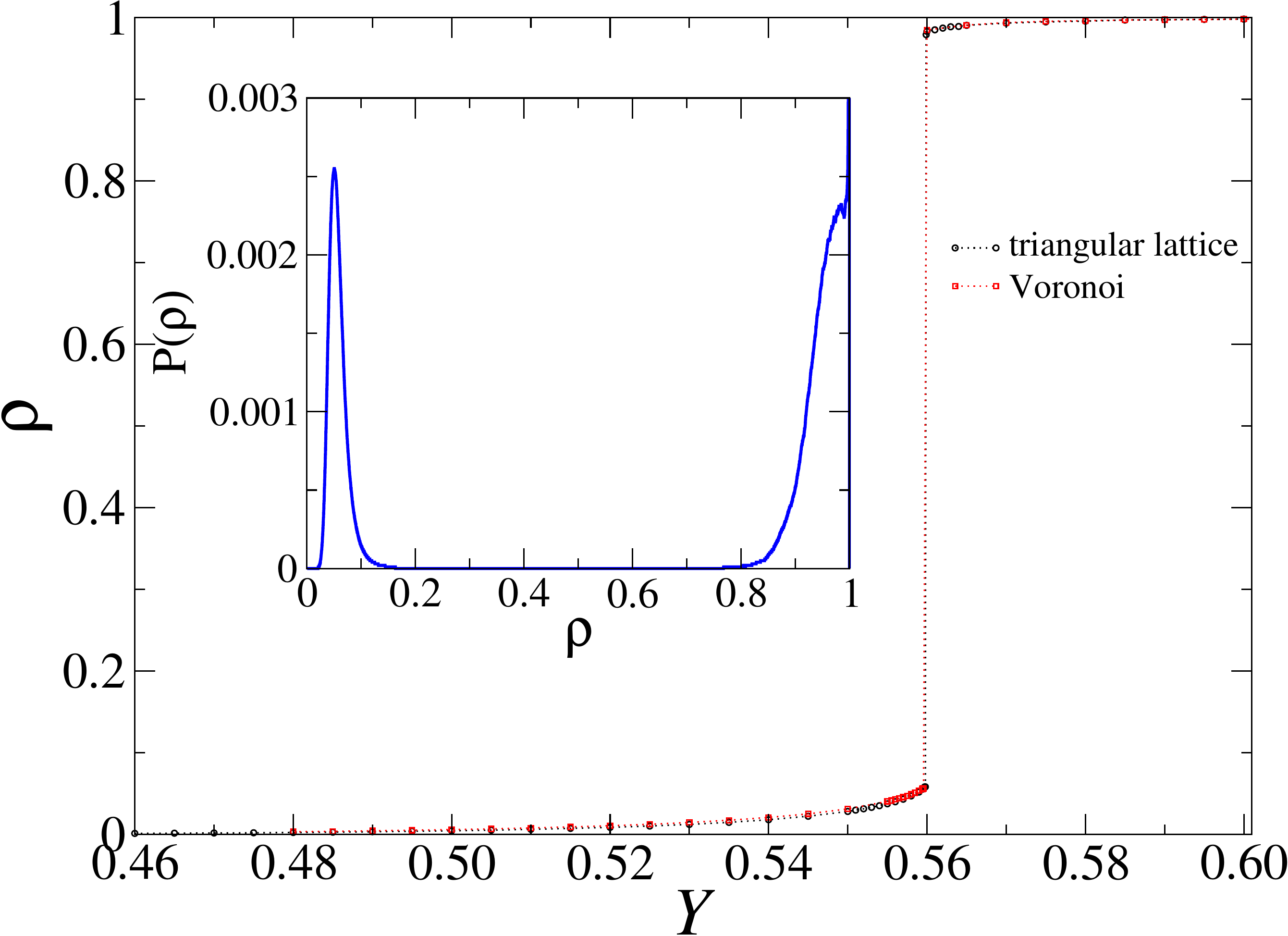}
\end{center}
\caption{(Color online) QS Density of CO sites in ZGB model on 
triangular and Voronoi lattices of linear system size $L=100$, showing a discontinuous
phase transition to the CO-poisoned absorbing state close to $Y=0.56$. Inset
shows the QS distribution for $Y=0.5590$ where the APT takes place.}
\label{zgbqs}
\end{figure}

Quasi-stationary analysis remains useful in the context of
discontinuous APTs~\cite{DeOliveira2015}. Considering the QS simulations we observe a
discontinuous phase transition from a low-density to a poisoned
(absorbing) CO state, as shown in Fig.~\ref{zgbqs} instead of a
rounded (continuous) transition expected for APTs in the presence of
relevant disorder~\cite{martin}. The inset of Fig.~\ref{zgbqs} shows
the QS probability distribution for the density of active sites near
the transition. We clearly observe a bimodal distribution, which is a
hallmark of discontinuous phase transition \cite{DeOliveira2015}.

\begin{figure}[ht]
\begin{center}
\includegraphics[width=8cm,clip=true]{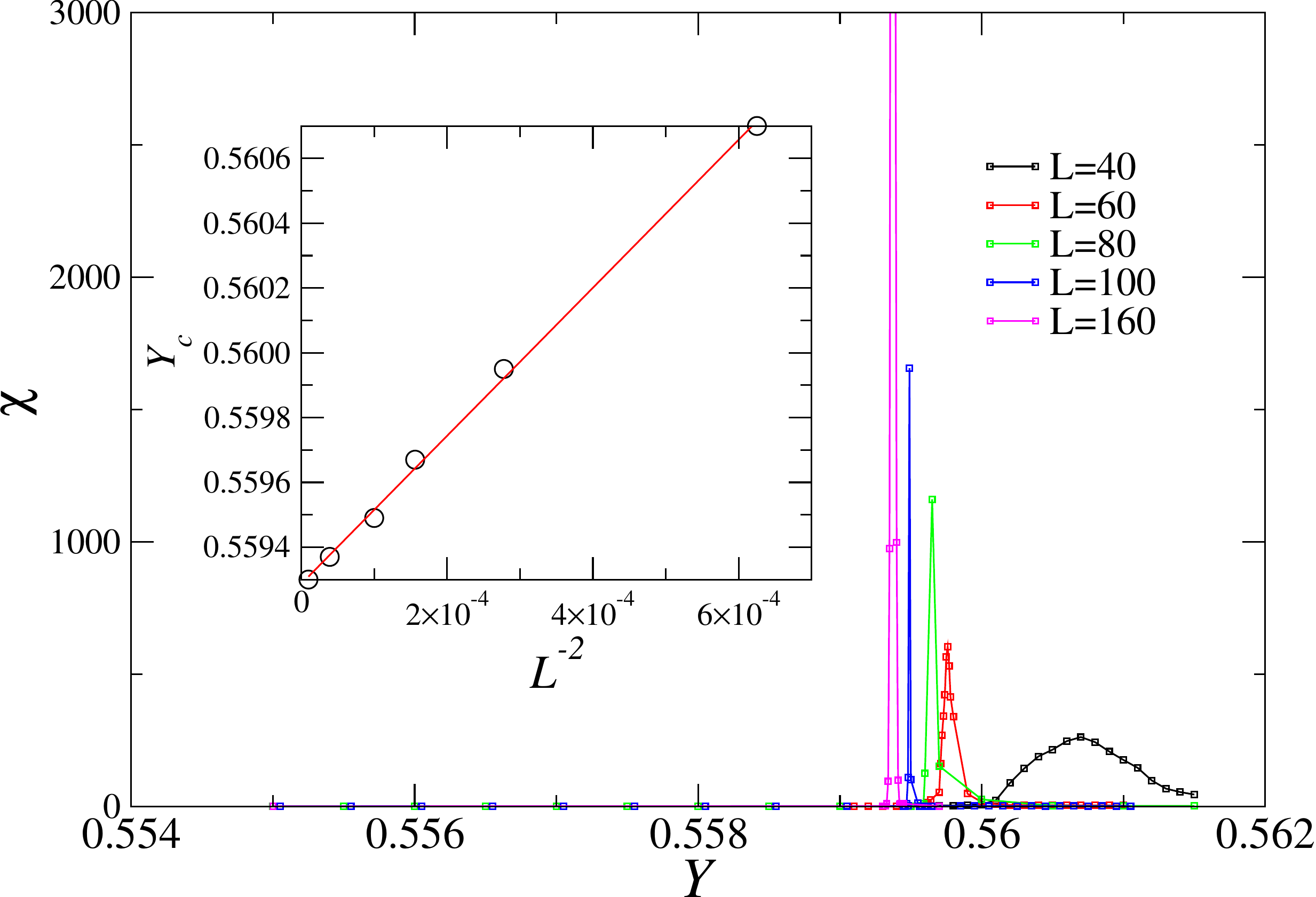}
\end{center}
\caption{(Color online) Quasi-stationary susceptibility on the ZGB model on VD lattices as a
function of $Y$ for different sizes. Inset: finite size scaling for the
susceptibility maxima in the range $L=40$ to $320$.}
\label{zgbvar}
\end{figure}

In analogy to equilibrium first-order phase transition, where at the
transition point a thermodynamical potential (such as the free energy)
is equal for both phases\cite{binder}, we can define the coexistence
value of the order parameter in which the area under the peaks of the
QS distribution related to each phase (active and absorbing) are equal
\cite{DeOliveira2015}. The intercept of the linear fit from this equal
histogram method yields $Y_c=0.55928(3)$. Such a value is very close
to the coexistence value $Y_c=0.5596(5)$ we found for the regular
triangular lattice (see inset of Fig.~\ref{zgbvar}).

The location of the maximum of  the susceptibility $\chi$, defined as
variance of the order parameter
$\chi=L^d(\lrangle{\rho^2}-\lrangle{\rho}^2)$, scales as $L^d$ in a
discontinuous APT \cite{AliSaif2009,Sinha2012,DeOliveira2015}.
Figure~\ref{zgbvar} shows the finite-size scaling of the transition
point which clearly scales inversely to the volume, confirming again
the disordered lattice does not alter the discontinuous nature of the 
transition.

\begin{figure}[htb]
\begin{center}
\includegraphics[width=8cm,clip=true]{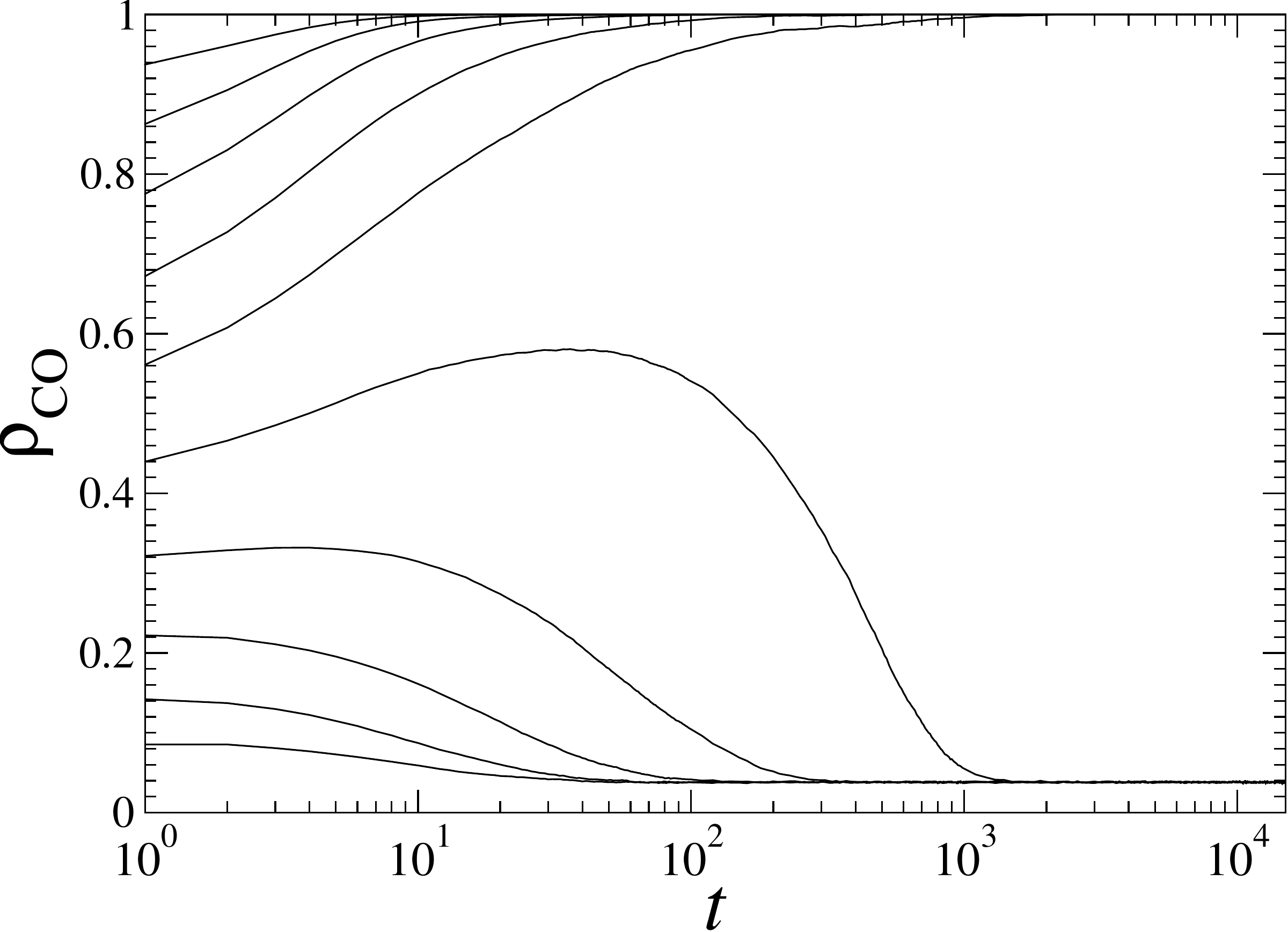}
\end{center}
\caption{
 Density of CO as a function of time for distinct initial conditions
 close to the transition point. Initial densities $\rho_{CO} = 0.0,
 ~0.1,~0.2,\cdots,0.9$, from bottom to top. Linear system size L =100 and
 $Y=0.5560$.}
\label{bistable}
\end{figure}

Further evidence of discontinuity of the phase transition
in presence of quenched coordination disorder, is
shown in Fig.~\ref{bistable}. Using conventional simulations, we observe the system bistability
around the transition point: depending on the initial density, a
homogeneous steady state may converge either to a stationary active
state of high CO$_2$ production (and small CO density) or to the
CO-poisoned (absorbing) state.
These results contrasts with those for uncorrelated disorder, for
which no matter its strength, the discontinuous transition is replaced
by a continuous one.

\subsection{Continuous APT}

The spreading analysis for the A model on VD lattices using the mean
number of active sites against time, with a single occupied site as
initial condition, provides a critical value $\lambda_c=0.322430(5)$,
which is smaller than $\lambda_c = 0.34047(1)$ found for the regular
triangular lattice with $q=6$. This difference is more significant
than that obtained for CP for these same lattices~\cite{oliveira2}
showing that effect of disorder in A model is stronger than in CP.
However, the critical behavior remains that of the clean system, exhibiting clear power laws with
spreading exponents very consistent with the DP class (results not shown).

\begin{figure}[h]
\begin{center}
\includegraphics[width=8cm,clip=true]{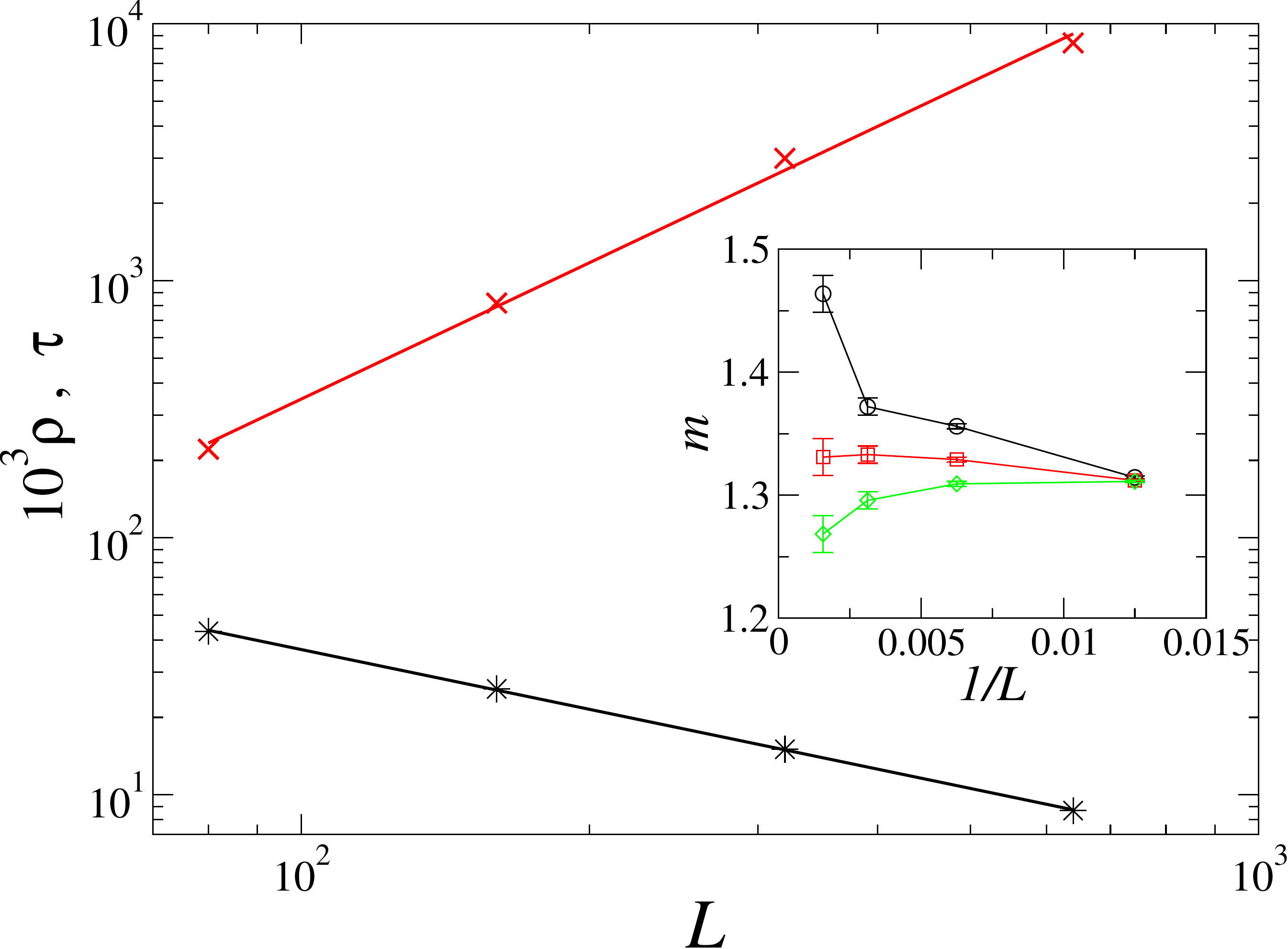}.
\end{center}
\caption{\label{fig:FSSVD} (Color online) FSS of the critical A model. Main:
Quasistationary density of active sites $\rho$ (stars) and lifetime of the
QS state $\tau$ (crosses) as a function of the system sizes $L$ for
$\lambda=0.32243$. Inset: Quasistationary moment ratio $m$ versus $1/L$,
for $\lambda = 0.32238$, $\lambda = 0.32242$, $\lambda = 0.32246$, from top
to bottom.}
\end{figure}

Figure~\ref{fig:FSSVD} shows that at the critical point we found the
QS density $\rho$ decays as a power law, $ \rho \sim
L^{-\beta/\nu_\perp}$, with $\beta/\nu_\perp=0.79(1)$. Besides, we
observe that the lifetime of the QS state also follows a power-law at
criticality, with $ \tau \sim L^{z}$, $z=1.73(5)$. Both values of the
exponents are close to the DP ones of $\beta/\nu_\perp=0.797(3)$ and
$z=1.7674(6)$~\cite{henkel08}. The inset of Fig.~\ref{fig:FSSVD} shows
the ratio $m=\lrangle{\rho^2}/\lrangle{\rho}$ around the criticality
for varying system sizes. From these data we found $m_c=1.33(1)$, in
agreement with the value $m_c = 1.3257(5)$ found for DP class in two
dimensions~\cite{dic-jaf}.  All results presented here confirm the
irrelevance of disorder of the VD lattice for the critical behavior of
the A model.

\begin{figure}[th]
\centering
\includegraphics[width=8cm]{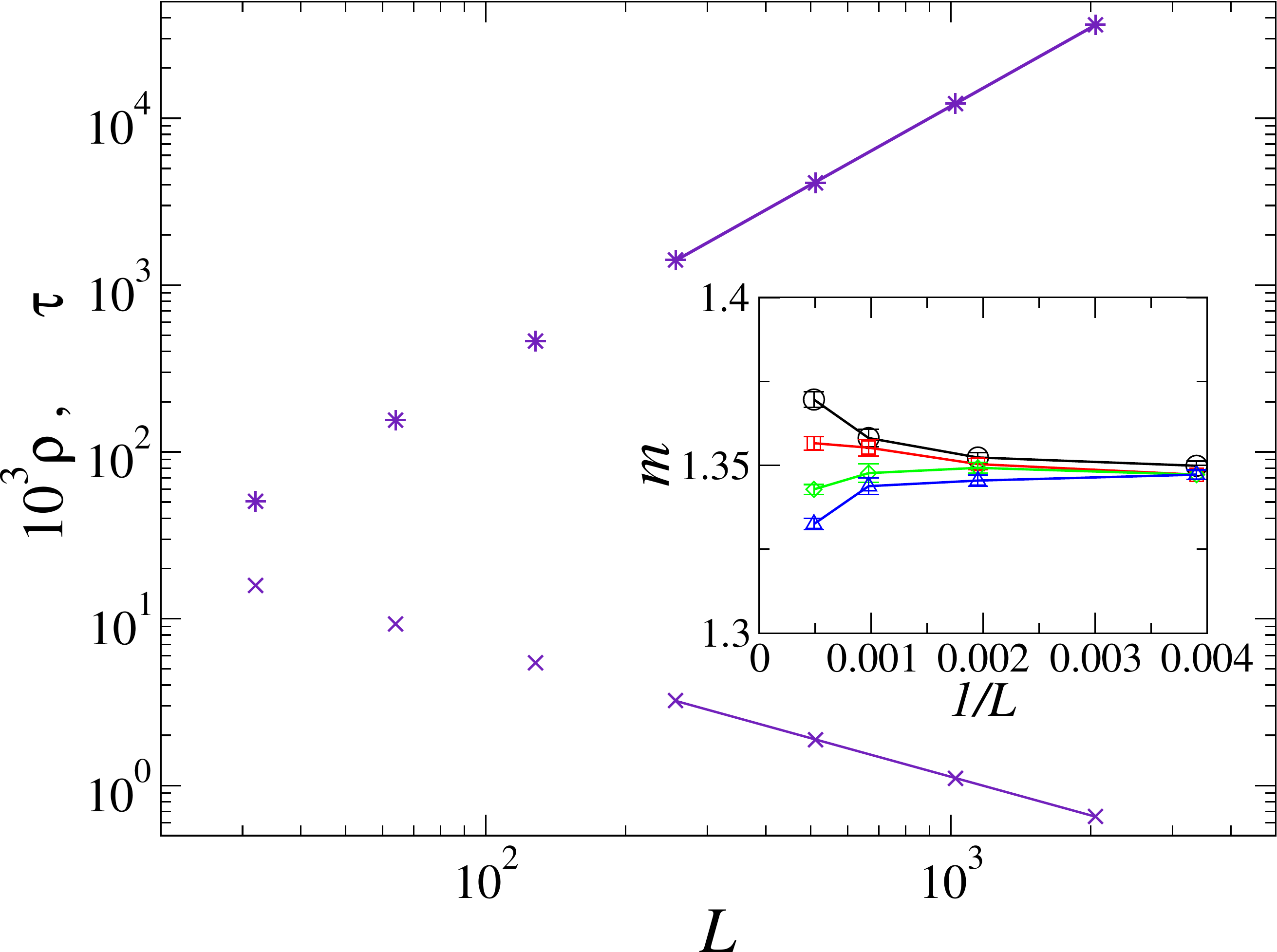}
\caption{(Color online) Critical Manna model on VD lattices. Main: Critical density of
active sites (crosses) and lifetime (stars) against lattice size. Inset:
Moment ratio $m=\lrangle{\rho^2}/\lrangle{\rho}^2$ against inverse of size
for $p=0.688800,~0.688805,~0.688810,~$and 0.688815 from top to bottom.}
\label{fig:Mannacrit}
\end{figure}

Lets now turn our attention to the Manna class. The correlation length
exponent $\nu_\perp=0.799$~\cite{Lubeck2002}, larger than  the DP value
0.7333, makes the modified Harris-Luck criterion
modified criterion $(d+1)\nu_\perp<2$ still not fulfilled for VD lattices~\cite{vojta14c}.
Critical point determination using moment ratios is shown in the inset of
Fig.~\ref{fig:Mannacrit} resulting in the estimate $p_c=0.688808(2)$ that
is smaller than the triangular lattice threshold~$p_c=0.69375(5)$. The
critical moment ratio is $m_c=1.35(1)$, which agrees with the value we
found for square lattices\footnote{Our estimate of $m$ does not agree with that of
Ref.~\cite{DaCunha2014} where a restricted version of the Manna model, in which
$N_i>2$ is forbidden, was considered.} $m_c=1.348(7)$ at the threshold $p_c=0.716957(2)$. 
The critical exponents we obtained
using $L\geqslant 256$ were $\beta/\nu_\perp=0.78(1)$ and
$\nu_\parallel/\nu_\perp=1.54(2)$ are also in striking agreement with the
Manna class exponents $\beta/\nu_\perp=0.80(2)$ and
$\nu_\parallel/\nu_\perp=1.53(5)$.

\begin{figure}
\centering
\includegraphics[width=8cm]{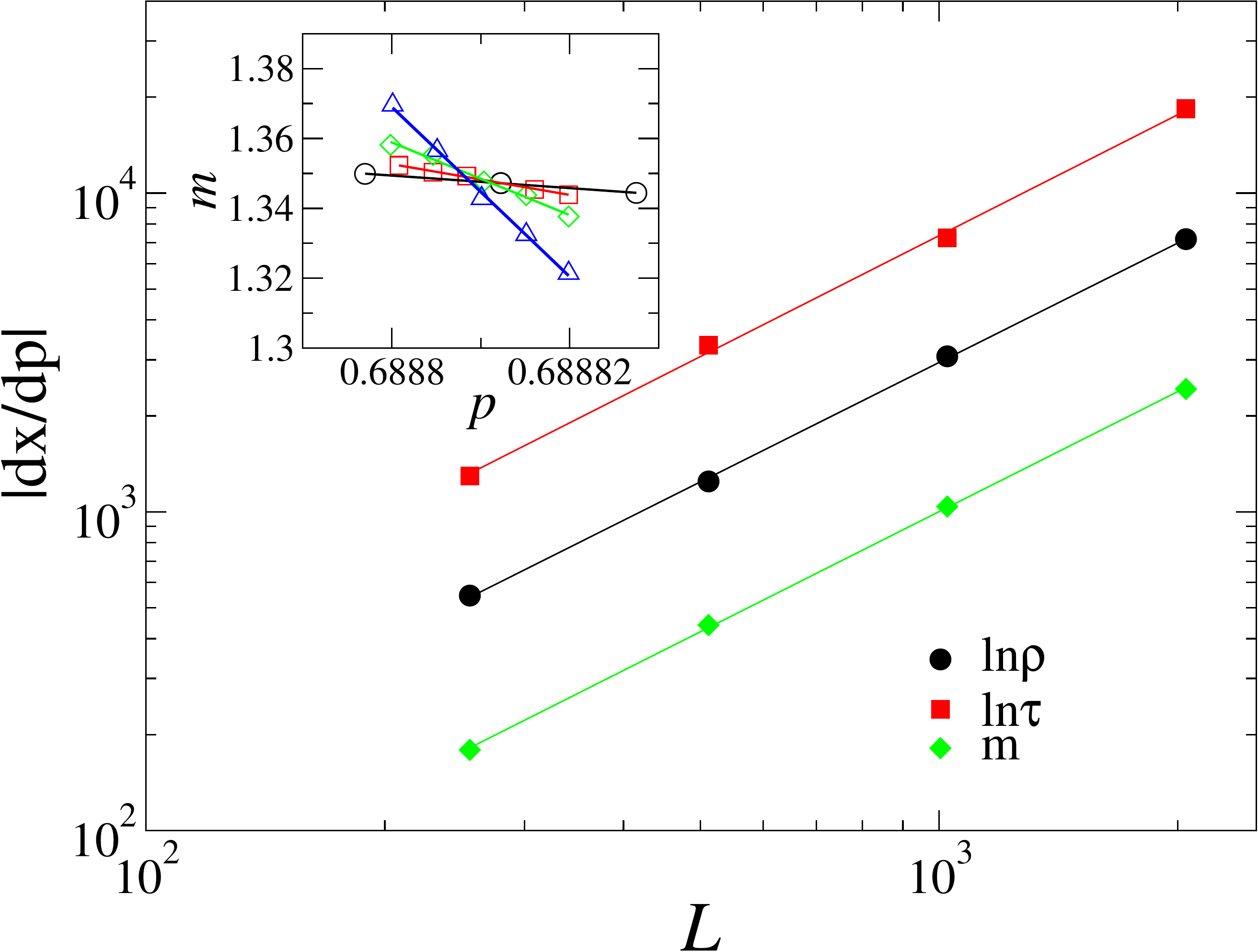}
\caption{(Color online) Determination of critical exponent $\nu_\perp$ for Manna model on VD lattices
using different quantities $x=\ln \rho,~\ln \tau$ and $m$. Inset: Moment
ratio against control parameter around the critical point for
$L=256$,~512,~1024, and 2048.}
\label{fig:dxdlb}
\end{figure}

It is known that critical  exponents  and moment ratios of the Manna
class in $d=2$ obtained via QS analysis are hardly distinguishable form DP
class~\cite{henkel08,Bonachela2007}. In order to provide a more incisive
verification that Manna model on VD lattice has exponents different from DP
we considered density around the critical point, which scales
as~\cite{Dickman2006}
\begin{equation}
\rho(\Delta,L)=\frac{1}{L^{\beta/\nu_\perp}}\mathcal{F}_\rho(L^{1/\nu_\perp} \Delta),
\end{equation}
where $\Delta=p-p_c$. This implies that
\begin{equation}
\left|\frac{\partial \ln \rho}{\partial p}\right| \sim L^{1/\nu_\perp}
\end{equation}
can be used to obtain the exponent $\nu_\perp$ explicitly. Similarly, for
the moment ratio we have $m(\Delta,L)=\mathcal{F}_m(L^{1/\nu_\perp}
\Delta)$ implying that $\nu_\perp$ can also be directly obtained from
\begin{equation}
\left|\frac{\partial m}{\partial p}\right| \sim L^{1/\nu_\perp}.
\end{equation}
A similar scaling law is expected for $\tau$. The inset of
Fig.\ref{fig:dxdlb} shows the moment ratios  around the
critical point where the slope clearly increase (in absolute values) with
size. The main plot shows the derivatives against size. Using
the three methods, we estimate  a critical exponent $1/\nu_\perp=1.252(10)$,
which is remarkably close to the exponent for Manna class 
$1/\nu_\perp=1.250(18)$~\cite{henkel08} and definitely ruling out the DP
value  $1/\nu_\perp=1.364(10)$~\cite{Marrobook}.

\section{Conclusions}

We investigate the effects of quenched coordination disorder in
continuous and discontinuous absorbing state phase transitions. 
Our extensive simulations of the ZGB model on the VD lattice reveal
the discontinuous nature of the absorbing state transition featured by
the model remains unchanged under such a kind of disorder.
Recently, it was shown that the Imry-Ma argument can be extended to
non-equilibrium situations including absorbing states, and in
addition, it was conjectured that first-order phase transitions cannot
appear in low-dimensional disordered systems with an absorbing state.
We showed that this is not always true: Our results for the ZGB model
raise the interesting possibility that disorder in nonequilibrium APT
may, under certain conditions, be irrelevant for the phase
coexistence. The underlying reason for this is that the fluctuations
induce correlated coordination  disorder exhibited by the VD Lattice
decay faster and are not able to preclude phase coexistence.

In the case of continuous APT, we performed large-scale simulations of the A
and Manna models on a Voronoi-Delaunay random lattice. Our results
confirm, as expected, that this kind of disorder does not alter the universality of
the continuous transitions, supporting that strong
anticorrelations present in the VD random lattice makes topological
disorder less relevant than uncorrelated randomness.

Our findings corroborate a recent work of Barghatti and
Vojta~\cite{vojta14c} which shows systematically that the disorder
fluctuations of the VD lattice are featured by strong anticorrelations
and decay faster than those of random uncorrelated disorder.
In particular, it was shown that the random VD lattice has wandering
exponent $\omega = 1/4$~\cite{vojta14c}. Hence, in this case, the
Harris-Luck criterion yields that random connectivity is irrelevant at
a clean critical point for $\nu_\perp > 2/3$ that is satisfied for
both Manna and DP universality classes. It is important to mention
that in contrast to the A model, which belongs to DP class, even the
strong disorder of uncorrelated lattice dilution (below the lattice
percolation threshold) was found to be irrelevant for Manna
class\cite{LeePRE2011,LeePRE2013,LeePRL}. Therefore, our results are consistent with
these findings, since the coordination disorder of VD lattice
is weaker than lattice dilution. In addition, we determined the exponent $1/\nu_\perp= 1.252(10)$
for Manna class on VD lattice  definitely ruling out the DP value  
$1/\nu_\perp= 1.364(10)$.

Further work should include the study of absorbing phase transitions
on a three-dimensional random VD lattice, since it does not belong to
the class of lattices with constrained total coordination
\cite{vojta14}. In particular, according to the Harris criterion, the
disorder might be relevant for the Manna class at least in three
dimensions and there might be a dimensional difference between two and
three dimensions. It would also be interesting to investigate if other
kinds of correlated disorder are irrelevant for phase coexistence. 

\begin{acknowledgments}


This work was supported by CNPq, CAPES and FAPEMIG, Brazil. M.M.O
thanks the kind hospitality at the Complex Systems and Statistical
Physics Group/University of Manchester, where part of this work was
done, and financial support from CAPES, under project BEX 10646/13-2.

\end{acknowledgments}



\begin{thebibliography}{66}%
\makeatletter
\providecommand \@ifxundefined [1]{%
 \@ifx{#1\undefined}
}%
\providecommand \@ifnum [1]{%
 \ifnum #1\expandafter \@firstoftwo
 \else \expandafter \@secondoftwo
 \fi
}%
\providecommand \@ifx [1]{%
 \ifx #1\expandafter \@firstoftwo
 \else \expandafter \@secondoftwo
 \fi
}%
\providecommand \natexlab [1]{#1}%
\providecommand \enquote  [1]{``#1''}%
\providecommand \bibnamefont  [1]{#1}%
\providecommand \bibfnamefont [1]{#1}%
\providecommand \citenamefont [1]{#1}%
\providecommand \href@noop [0]{\@secondoftwo}%
\providecommand \href [0]{\begingroup \@sanitize@url \@href}%
\providecommand \@href[1]{\@@startlink{#1}\@@href}%
\providecommand \@@href[1]{\endgroup#1\@@endlink}%
\providecommand \@sanitize@url [0]{\catcode `\\12\catcode `\$12\catcode
  `\&12\catcode `\#12\catcode `\^12\catcode `\_12\catcode `\%12\relax}%
\providecommand \@@startlink[1]{}%
\providecommand \@@endlink[0]{}%
\providecommand \url  [0]{\begingroup\@sanitize@url \@url }%
\providecommand \@url [1]{\endgroup\@href {#1}{\urlprefix }}%
\providecommand \urlprefix  [0]{URL }%
\providecommand \Eprint [0]{\href }%
\providecommand \doibase [0]{http://dx.doi.org/}%
\providecommand \selectlanguage [0]{\@gobble}%
\providecommand \bibinfo  [0]{\@secondoftwo}%
\providecommand \bibfield  [0]{\@secondoftwo}%
\providecommand \translation [1]{[#1]}%
\providecommand \BibitemOpen [0]{}%
\providecommand \bibitemStop [0]{}%
\providecommand \bibitemNoStop [0]{.\EOS\space}%
\providecommand \EOS [0]{\spacefactor3000\relax}%
\providecommand \BibitemShut  [1]{\csname bibitem#1\endcsname}%
\let\auto@bib@innerbib\@empty
\bibitem [{\citenamefont {Marro}\ and\ \citenamefont
  {Dickman}(1999)}]{Marrobook}%
  \BibitemOpen
  \bibfield  {author} {\bibinfo {author} {\bibfnamefont {J.}~\bibnamefont
  {Marro}}\ and\ \bibinfo {author} {\bibfnamefont {R.}~\bibnamefont
  {Dickman}},\ }\href@noop {} {\emph {\bibinfo {title} {Nonequilibrium phase
  transitions in lattice models}}}\ (\bibinfo  {publisher} {Cambridge
  University Press},\ \bibinfo {address} {Cambridge},\ \bibinfo {year}
  {1999})\BibitemShut {NoStop}%
\bibitem [{\citenamefont {Henkel}\ \emph {et~al.}(2008)\citenamefont {Henkel},
  \citenamefont {Hinrichsen}, \citenamefont {L{\"u}},\ and\ \citenamefont
  {Pleimling}}]{henkel08}%
  \BibitemOpen
  \bibfield  {author} {\bibinfo {author} {\bibfnamefont {M.}~\bibnamefont
  {Henkel}}, \bibinfo {author} {\bibfnamefont {H.}~\bibnamefont {Hinrichsen}},
  \bibinfo {author} {\bibfnamefont {S.}~\bibnamefont {L{\"u}}}, \ and\ \bibinfo
  {author} {\bibfnamefont {M.}~\bibnamefont {Pleimling}},\ }\href@noop {}
  {\emph {\bibinfo {title} {Non-equilibrium phase transitions}}},\
  Vol.~\bibinfo {volume} {1}\ (\bibinfo  {publisher} {Springer},\ \bibinfo
  {address} {Dordrecht, Netherlands},\ \bibinfo {year} {2008})\BibitemShut
  {NoStop}%
\bibitem [{\citenamefont {\'Odor}(2004)}]{odor04}%
  \BibitemOpen
  \bibfield  {author} {\bibinfo {author} {\bibfnamefont {G.}~\bibnamefont
  {\'Odor}},\ }\href@noop {} {\bibfield  {journal} {\bibinfo  {journal} {Rev.
  Mod. Phys.}\ }\textbf {\bibinfo {volume} {76}},\ \bibinfo {pages} {663}
  (\bibinfo {year} {2004})}\BibitemShut {NoStop}%
\bibitem [{\citenamefont {Ziff}\ \emph {et~al.}(1986)\citenamefont {Ziff},
  \citenamefont {Gulari},\ and\ \citenamefont {Barshad}}]{zgb}%
  \BibitemOpen
  \bibfield  {author} {\bibinfo {author} {\bibfnamefont {R.~M.}\ \bibnamefont
  {Ziff}}, \bibinfo {author} {\bibfnamefont {E.}~\bibnamefont {Gulari}}, \ and\
  \bibinfo {author} {\bibfnamefont {Y.}~\bibnamefont {Barshad}},\ }\href@noop
  {} {\bibfield  {journal} {\bibinfo  {journal} {Phys. Rev. Lett.}\ }\textbf
  {\bibinfo {volume} {56}},\ \bibinfo {pages} {2553} (\bibinfo {year}
  {1986})}\BibitemShut {NoStop}%
\bibitem [{\citenamefont {Tang}\ and\ \citenamefont {Leschhorn}(1992)}]{tang}%
  \BibitemOpen
  \bibfield  {author} {\bibinfo {author} {\bibfnamefont {L.-H.}\ \bibnamefont
  {Tang}}\ and\ \bibinfo {author} {\bibfnamefont {H.}~\bibnamefont
  {Leschhorn}},\ }\href@noop {} {\bibfield  {journal} {\bibinfo  {journal}
  {Phys. Rev. A}\ }\textbf {\bibinfo {volume} {45}},\ \bibinfo {pages} {R8309}
  (\bibinfo {year} {1992})}\BibitemShut {NoStop}%
\bibitem [{\citenamefont {Pastor-Satorras}\ \emph {et~al.}(2015)\citenamefont
  {Pastor-Satorras}, \citenamefont {Castellano}, \citenamefont {Van~Mieghem},\
  and\ \citenamefont {Vespignani}}]{pastor2014}%
  \BibitemOpen
  \bibfield  {author} {\bibinfo {author} {\bibfnamefont {R.}~\bibnamefont
  {Pastor-Satorras}}, \bibinfo {author} {\bibfnamefont {C.}~\bibnamefont
  {Castellano}}, \bibinfo {author} {\bibfnamefont {P.}~\bibnamefont
  {Van~Mieghem}}, \ and\ \bibinfo {author} {\bibfnamefont {A.}~\bibnamefont
  {Vespignani}},\ }\href {\doibase 10.1103/RevModPhys.87.925} {\bibfield
  {journal} {\bibinfo  {journal} {Rev. Mod. Phys.}\ }\textbf {\bibinfo {volume}
  {87}},\ \bibinfo {pages} {925} (\bibinfo {year} {2015})}\BibitemShut
  {NoStop}%
\bibitem [{\citenamefont {Takeuchi}\ \emph {et~al.}(2007)\citenamefont
  {Takeuchi}, \citenamefont {Kuroda}, \citenamefont {Chat\'e},\ and\
  \citenamefont {Sano}}]{take07}%
  \BibitemOpen
  \bibfield  {author} {\bibinfo {author} {\bibfnamefont {K.~A.}\ \bibnamefont
  {Takeuchi}}, \bibinfo {author} {\bibfnamefont {M.}~\bibnamefont {Kuroda}},
  \bibinfo {author} {\bibfnamefont {H.}~\bibnamefont {Chat\'e}}, \ and\
  \bibinfo {author} {\bibfnamefont {M.}~\bibnamefont {Sano}},\ }\href {\doibase
  10.1103/PhysRevLett.99.234503} {\bibfield  {journal} {\bibinfo  {journal}
  {Phys. Rev. Lett.}\ }\textbf {\bibinfo {volume} {99}},\ \bibinfo {pages}
  {234503} (\bibinfo {year} {2007})}\BibitemShut {NoStop}%
\bibitem [{\citenamefont {Laurent}\ \emph {et~al.}(2008)\citenamefont
  {Laurent}, \citenamefont {Chaikin}, \citenamefont {P.},\ and\ \citenamefont
  {Pine}}]{pine}%
  \BibitemOpen
  \bibfield  {author} {\bibinfo {author} {\bibfnamefont {C.}~\bibnamefont
  {Laurent}}, \bibinfo {author} {\bibnamefont {Chaikin}}, \bibinfo {author}
  {\bibfnamefont {P.~M. G.~J.}\ \bibnamefont {P.}}, \ and\ \bibinfo {author}
  {\bibfnamefont {D.~J.}\ \bibnamefont {Pine}},\ }\href@noop {} {\bibfield
  {journal} {\bibinfo  {journal} {Nat. Phys.}\ }\textbf {\bibinfo {volume}
  {4}},\ \bibinfo {pages} {420,424} (\bibinfo {year} {2008})}\BibitemShut
  {NoStop}%
\bibitem [{\citenamefont {Okuma}\ \emph {et~al.}(2011)\citenamefont {Okuma},
  \citenamefont {Tsugawa},\ and\ \citenamefont {Motohashi}}]{okuma}%
  \BibitemOpen
  \bibfield  {author} {\bibinfo {author} {\bibfnamefont {S.}~\bibnamefont
  {Okuma}}, \bibinfo {author} {\bibfnamefont {Y.}~\bibnamefont {Tsugawa}}, \
  and\ \bibinfo {author} {\bibfnamefont {A.}~\bibnamefont {Motohashi}},\ }\href
  {\doibase 10.1103/PhysRevB.83.012503} {\bibfield  {journal} {\bibinfo
  {journal} {Phys. Rev. B}\ }\textbf {\bibinfo {volume} {83}},\ \bibinfo
  {pages} {012503} (\bibinfo {year} {2011})}\BibitemShut {NoStop}%
\bibitem [{\citenamefont {Janssen}(1981)}]{gras}%
  \BibitemOpen
  \bibfield  {author} {\bibinfo {author} {\bibfnamefont {H.}~\bibnamefont
  {Janssen}},\ }\href@noop {} {\bibfield  {journal} {\bibinfo  {journal} {Z
  Phys. B}\ }\textbf {\bibinfo {volume} {42}},\ \bibinfo {pages} {151}
  (\bibinfo {year} {1981})}\BibitemShut {NoStop}%
\bibitem [{\citenamefont {Grassberger}(1982)}]{jans}%
  \BibitemOpen
  \bibfield  {author} {\bibinfo {author} {\bibfnamefont {P.}~\bibnamefont
  {Grassberger}},\ }\href@noop {} {\bibfield  {journal} {\bibinfo  {journal} {Z
  Phys. B}\ }\textbf {\bibinfo {volume} {47}},\ \bibinfo {pages} {365}
  (\bibinfo {year} {1982})}\BibitemShut {NoStop}%
\bibitem [{\citenamefont {Hinrichsen}(2000)}]{hinri00b}%
  \BibitemOpen
  \bibfield  {author} {\bibinfo {author} {\bibfnamefont {H.}~\bibnamefont
  {Hinrichsen}},\ }\href@noop {} {\bibfield  {journal} {\bibinfo  {journal}
  {Braz. J. Phys.}\ }\textbf {\bibinfo {volume} {30}},\ \bibinfo {pages} {69 }
  (\bibinfo {year} {2000})}\BibitemShut {NoStop}%
\bibitem [{\citenamefont {Noest}(1986)}]{noestPRL}%
  \BibitemOpen
  \bibfield  {author} {\bibinfo {author} {\bibfnamefont {A.~J.}\ \bibnamefont
  {Noest}},\ }\href {\doibase 10.1103/PhysRevLett.57.90} {\bibfield  {journal}
  {\bibinfo  {journal} {Phys. Rev. Lett.}\ }\textbf {\bibinfo {volume} {57}},\
  \bibinfo {pages} {90} (\bibinfo {year} {1986})}\BibitemShut {NoStop}%
\bibitem [{\citenamefont {Noest}(1988)}]{noestPRB}%
  \BibitemOpen
  \bibfield  {author} {\bibinfo {author} {\bibfnamefont {A.~J.}\ \bibnamefont
  {Noest}},\ }\href {\doibase 10.1103/PhysRevB.38.2715} {\bibfield  {journal}
  {\bibinfo  {journal} {Phys. Rev. B}\ }\textbf {\bibinfo {volume} {38}},\
  \bibinfo {pages} {2715} (\bibinfo {year} {1988})}\BibitemShut {NoStop}%
\bibitem [{\citenamefont {Moreira}\ and\ \citenamefont
  {Dickman}(1996)}]{adr-dic96}%
  \BibitemOpen
  \bibfield  {author} {\bibinfo {author} {\bibfnamefont {A.~G.}\ \bibnamefont
  {Moreira}}\ and\ \bibinfo {author} {\bibfnamefont {R.}~\bibnamefont
  {Dickman}},\ }\href {\doibase 10.1103/PhysRevE.54.R3090} {\bibfield
  {journal} {\bibinfo  {journal} {Phys. Rev. E}\ }\textbf {\bibinfo {volume}
  {54}},\ \bibinfo {pages} {R3090} (\bibinfo {year} {1996})}\BibitemShut
  {NoStop}%
\bibitem [{\citenamefont {Dickman}\ and\ \citenamefont
  {Moreira}(1998)}]{adr-dic98}%
  \BibitemOpen
  \bibfield  {author} {\bibinfo {author} {\bibfnamefont {R.}~\bibnamefont
  {Dickman}}\ and\ \bibinfo {author} {\bibfnamefont {A.~G.}\ \bibnamefont
  {Moreira}},\ }\href {\doibase 10.1103/PhysRevE.57.1263} {\bibfield  {journal}
  {\bibinfo  {journal} {Phys. Rev. E}\ }\textbf {\bibinfo {volume} {57}},\
  \bibinfo {pages} {1263} (\bibinfo {year} {1998})}\BibitemShut {NoStop}%
\bibitem [{\citenamefont {Vojta}\ and\ \citenamefont {Lee}(2006)}]{vojta06}%
  \BibitemOpen
  \bibfield  {author} {\bibinfo {author} {\bibfnamefont {T.}~\bibnamefont
  {Vojta}}\ and\ \bibinfo {author} {\bibfnamefont {M.~Y.}\ \bibnamefont
  {Lee}},\ }\href {\doibase 10.1103/PhysRevLett.96.035701} {\bibfield
  {journal} {\bibinfo  {journal} {Phys. Rev. Lett.}\ }\textbf {\bibinfo
  {volume} {96}},\ \bibinfo {pages} {035701} (\bibinfo {year}
  {2006})}\BibitemShut {NoStop}%
\bibitem [{\citenamefont {de~Oliveira}\ and\ \citenamefont
  {Ferreira}(2008)}]{DeOliveira2008}%
  \BibitemOpen
  \bibfield  {author} {\bibinfo {author} {\bibfnamefont {M.~M.}\ \bibnamefont
  {de~Oliveira}}\ and\ \bibinfo {author} {\bibfnamefont {S.~C.}\ \bibnamefont
  {Ferreira}},\ }\href@noop {} {\bibfield  {journal} {\bibinfo  {journal} {J.
  Stat. Mech.: Theor. Exp.}\ }\textbf {\bibinfo {volume} {2008}},\ \bibinfo
  {pages} {P11001} (\bibinfo {year} {2008})}\BibitemShut {NoStop}%
\bibitem [{\citenamefont {Bramson}\ \emph {et~al.}(1991)\citenamefont
  {Bramson}, \citenamefont {Durrett},\ and\ \citenamefont
  {Schonmann}}]{durrett}%
  \BibitemOpen
  \bibfield  {author} {\bibinfo {author} {\bibfnamefont {M.}~\bibnamefont
  {Bramson}}, \bibinfo {author} {\bibfnamefont {R.}~\bibnamefont {Durrett}}, \
  and\ \bibinfo {author} {\bibfnamefont {R.~H.}\ \bibnamefont {Schonmann}},\
  }\href@noop {} {\bibfield  {journal} {\bibinfo  {journal} {Ann. Probab.}\
  }\textbf {\bibinfo {volume} {19}},\ \bibinfo {pages} {pp. 960} (\bibinfo
  {year} {1991})}\BibitemShut {NoStop}%
\bibitem [{\citenamefont {Faria}\ \emph {et~al.}(2008)\citenamefont {Faria},
  \citenamefont {Ribeiro},\ and\ \citenamefont {S.~A.~Salinas}}]{salinas08}%
  \BibitemOpen
  \bibfield  {author} {\bibinfo {author} {\bibfnamefont {M.~S.}\ \bibnamefont
  {Faria}}, \bibinfo {author} {\bibfnamefont {D.~J.}\ \bibnamefont {Ribeiro}},
  \ and\ \bibinfo {author} {\bibfnamefont {J.}~\bibnamefont {S.~A.~Salinas}},\
  }\href@noop {} {\bibfield  {journal} {\bibinfo  {journal} {J. Stat. Mech.:
  Theor. Exp.}\ ,\ \bibinfo {pages} {P11001}} (\bibinfo {year}
  {2008})}\BibitemShut {NoStop}%
\bibitem [{\citenamefont {Barghathi}\ \emph {et~al.}(2014)\citenamefont
  {Barghathi}, \citenamefont {Nozadze},\ and\ \citenamefont
  {Vojta}}]{vojta14b}%
  \BibitemOpen
  \bibfield  {author} {\bibinfo {author} {\bibfnamefont {H.}~\bibnamefont
  {Barghathi}}, \bibinfo {author} {\bibfnamefont {D.}~\bibnamefont {Nozadze}},
  \ and\ \bibinfo {author} {\bibfnamefont {T.}~\bibnamefont {Vojta}},\ }\href
  {\doibase 10.1103/PhysRevE.89.012112} {\bibfield  {journal} {\bibinfo
  {journal} {Phys. Rev. E}\ }\textbf {\bibinfo {volume} {89}},\ \bibinfo
  {pages} {012112} (\bibinfo {year} {2014})}\BibitemShut {NoStop}%
\bibitem [{\citenamefont {Vojta}(2006)}]{vojta06b}%
  \BibitemOpen
  \bibfield  {author} {\bibinfo {author} {\bibfnamefont {T.}~\bibnamefont
  {Vojta}},\ }\href@noop {} {\bibfield  {journal} {\bibinfo  {journal} {J.
  Phys. A: Math. Gen.}\ }\textbf {\bibinfo {volume} {39}},\ \bibinfo {pages}
  {R143} (\bibinfo {year} {2006})}\BibitemShut {NoStop}%
\bibitem [{\citenamefont {Hooyberghs}\ \emph {et~al.}(2003)\citenamefont
  {Hooyberghs}, \citenamefont {Igl\'oi},\ and\ \citenamefont
  {Vanderzande}}]{HooyberghsPRL}%
  \BibitemOpen
  \bibfield  {author} {\bibinfo {author} {\bibfnamefont {J.}~\bibnamefont
  {Hooyberghs}}, \bibinfo {author} {\bibfnamefont {F.}~\bibnamefont {Igl\'oi}},
  \ and\ \bibinfo {author} {\bibfnamefont {C.}~\bibnamefont {Vanderzande}},\
  }\href@noop {} {\bibfield  {journal} {\bibinfo  {journal} {Phys. Rev. Lett.}\
  }\textbf {\bibinfo {volume} {90}},\ \bibinfo {pages} {100601} (\bibinfo
  {year} {2003})}\BibitemShut {NoStop}%
\bibitem [{\citenamefont {Hooyberghs}\ \emph {et~al.}(2004)\citenamefont
  {Hooyberghs}, \citenamefont {Igl\'oi},\ and\ \citenamefont
  {Vanderzande}}]{HooyberghsPRE}%
  \BibitemOpen
  \bibfield  {author} {\bibinfo {author} {\bibfnamefont {J.}~\bibnamefont
  {Hooyberghs}}, \bibinfo {author} {\bibfnamefont {F.}~\bibnamefont {Igl\'oi}},
  \ and\ \bibinfo {author} {\bibfnamefont {C.}~\bibnamefont {Vanderzande}},\
  }\href@noop {} {\bibfield  {journal} {\bibinfo  {journal} {Phys. Rev. E}\
  }\textbf {\bibinfo {volume} {69}},\ \bibinfo {pages} {066140} (\bibinfo
  {year} {2004})}\BibitemShut {NoStop}%
\bibitem [{\citenamefont {Vojta}\ and\ \citenamefont
  {Dickison}(2005)}]{vojta05}%
  \BibitemOpen
  \bibfield  {author} {\bibinfo {author} {\bibfnamefont {T.}~\bibnamefont
  {Vojta}}\ and\ \bibinfo {author} {\bibfnamefont {M.}~\bibnamefont
  {Dickison}},\ }\href@noop {} {\bibfield  {journal} {\bibinfo  {journal}
  {Phys. Rev. E}\ }\textbf {\bibinfo {volume} {72}},\ \bibinfo {pages} {036126}
  (\bibinfo {year} {2005})}\BibitemShut {NoStop}%
\bibitem [{\citenamefont {Vojta}\ \emph {et~al.}(2009)\citenamefont {Vojta},
  \citenamefont {Farquhar},\ and\ \citenamefont {Mast}}]{vojta09}%
  \BibitemOpen
  \bibfield  {author} {\bibinfo {author} {\bibfnamefont {T.}~\bibnamefont
  {Vojta}}, \bibinfo {author} {\bibfnamefont {A.}~\bibnamefont {Farquhar}}, \
  and\ \bibinfo {author} {\bibfnamefont {J.}~\bibnamefont {Mast}},\ }\href
  {\doibase 10.1103/PhysRevE.79.011111} {\bibfield  {journal} {\bibinfo
  {journal} {Phys. Rev. E}\ }\textbf {\bibinfo {volume} {79}},\ \bibinfo
  {pages} {011111} (\bibinfo {year} {2009})}\BibitemShut {NoStop}%
\bibitem [{\citenamefont {Harris}(1974{\natexlab{a}})}]{harris74}%
  \BibitemOpen
  \bibfield  {author} {\bibinfo {author} {\bibfnamefont {A.~B.}\ \bibnamefont
  {Harris}},\ }\href@noop {} {\bibfield  {journal} {\bibinfo  {journal} {J.
  Phys. C: Solid State Physics}\ }\textbf {\bibinfo {volume} {7}},\ \bibinfo
  {pages} {1671} (\bibinfo {year} {1974}{\natexlab{a}})}\BibitemShut {NoStop}%
\bibitem [{\citenamefont {Jensen}(1992)}]{jensen92}%
  \BibitemOpen
  \bibfield  {author} {\bibinfo {author} {\bibfnamefont {I.}~\bibnamefont
  {Jensen}},\ }\href {\doibase 10.1103/PhysRevA.45.R563} {\bibfield  {journal}
  {\bibinfo  {journal} {Phys. Rev. A}\ }\textbf {\bibinfo {volume} {45}},\
  \bibinfo {pages} {R563} (\bibinfo {year} {1992})}\BibitemShut {NoStop}%
\bibitem [{\citenamefont {Jensen}(1999)}]{jensen99}%
  \BibitemOpen
  \bibfield  {author} {\bibinfo {author} {\bibfnamefont {I.}~\bibnamefont
  {Jensen}},\ }\href@noop {} {\bibfield  {journal} {\bibinfo  {journal} {J.
  Phys. A: Math. Gen.}\ }\textbf {\bibinfo {volume} {32}},\ \bibinfo {pages}
  {5233} (\bibinfo {year} {1999})}\BibitemShut {NoStop}%
\bibitem [{\citenamefont {Voigt}\ and\ \citenamefont {Ziff}(1997)}]{voigt97}%
  \BibitemOpen
  \bibfield  {author} {\bibinfo {author} {\bibfnamefont {C.~A.}\ \bibnamefont
  {Voigt}}\ and\ \bibinfo {author} {\bibfnamefont {R.~M.}\ \bibnamefont
  {Ziff}},\ }\href {\doibase 10.1103/PhysRevE.56.R6241} {\bibfield  {journal}
  {\bibinfo  {journal} {Phys. Rev. E}\ }\textbf {\bibinfo {volume} {56}},\
  \bibinfo {pages} {R6241} (\bibinfo {year} {1997})}\BibitemShut {NoStop}%
\bibitem [{\citenamefont {Manna}(1991)}]{manna}%
  \BibitemOpen
  \bibfield  {author} {\bibinfo {author} {\bibfnamefont {S.~S.}\ \bibnamefont
  {Manna}},\ }\href@noop {} {\bibfield  {journal} {\bibinfo  {journal} {J.
  Phys. A: Math. Gen.}\ }\textbf {\bibinfo {volume} {24}},\ \bibinfo {pages}
  {L363} (\bibinfo {year} {1991})}\BibitemShut {NoStop}%
\bibitem [{\citenamefont {Lee}(2011)}]{LeePRE2011}%
  \BibitemOpen
  \bibfield  {author} {\bibinfo {author} {\bibfnamefont {S.~B.}\ \bibnamefont
  {Lee}},\ }\href@noop {} {\bibfield  {journal} {\bibinfo  {journal} {Phys.
  Rev. E}\ }\textbf {\bibinfo {volume} {84}},\ \bibinfo {pages} {041123}
  (\bibinfo {year} {2011})}\BibitemShut {NoStop}%
\bibitem [{\citenamefont {Lee}\ and\ \citenamefont {Kim}(2013)}]{LeePRE2013}%
  \BibitemOpen
  \bibfield  {author} {\bibinfo {author} {\bibfnamefont {S.~B.}\ \bibnamefont
  {Lee}}\ and\ \bibinfo {author} {\bibfnamefont {J.~S.}\ \bibnamefont {Kim}},\
  }\href@noop {} {\bibfield  {journal} {\bibinfo  {journal} {Phys. Rev. E}\
  }\textbf {\bibinfo {volume} {87}},\ \bibinfo {pages} {032117} (\bibinfo
  {year} {2013})}\BibitemShut {NoStop}%
\bibitem [{\citenamefont {Lee}(2013)}]{LeePRL}%
  \BibitemOpen
  \bibfield  {author} {\bibinfo {author} {\bibfnamefont {S.~B.}\ \bibnamefont
  {Lee}},\ }\href@noop {} {\bibfield  {journal} {\bibinfo  {journal} {Phys.
  Rev. Lett.}\ }\textbf {\bibinfo {volume} {110}},\ \bibinfo {pages} {159601}
  (\bibinfo {year} {2013})}\BibitemShut {NoStop}%
\bibitem [{\citenamefont {Imry}\ and\ \citenamefont {Ma}(1975)}]{ma}%
  \BibitemOpen
  \bibfield  {author} {\bibinfo {author} {\bibfnamefont {Y.}~\bibnamefont
  {Imry}}\ and\ \bibinfo {author} {\bibfnamefont {S.-k.}\ \bibnamefont {Ma}},\
  }\href {\doibase 10.1103/PhysRevLett.35.1399} {\bibfield  {journal} {\bibinfo
   {journal} {Phys. Rev. Lett.}\ }\textbf {\bibinfo {volume} {35}},\ \bibinfo
  {pages} {1399} (\bibinfo {year} {1975})}\BibitemShut {NoStop}%
\bibitem [{\citenamefont {Hui}\ and\ \citenamefont {Berker}(1989)}]{ma2}%
  \BibitemOpen
  \bibfield  {author} {\bibinfo {author} {\bibfnamefont {K.}~\bibnamefont
  {Hui}}\ and\ \bibinfo {author} {\bibfnamefont {A.~N.}\ \bibnamefont
  {Berker}},\ }\href {\doibase 10.1103/PhysRevLett.62.2507} {\bibfield
  {journal} {\bibinfo  {journal} {Phys. Rev. Lett.}\ }\textbf {\bibinfo
  {volume} {62}},\ \bibinfo {pages} {2507} (\bibinfo {year}
  {1989})}\BibitemShut {NoStop}%
\bibitem [{\citenamefont {Buend\'{\i}a}\ and\ \citenamefont
  {Rikvold}(2012)}]{buendia}%
  \BibitemOpen
  \bibfield  {author} {\bibinfo {author} {\bibfnamefont {G.~M.}\ \bibnamefont
  {Buend\'{\i}a}}\ and\ \bibinfo {author} {\bibfnamefont {P.~A.}\ \bibnamefont
  {Rikvold}},\ }\href@noop {} {\bibfield  {journal} {\bibinfo  {journal} {Phys.
  Rev. E}\ }\textbf {\bibinfo {volume} {85}},\ \bibinfo {pages} {031143}
  (\bibinfo {year} {2012})}\BibitemShut {NoStop}%
\bibitem [{\citenamefont {{Buend\'{\i}a}}\ and\ \citenamefont
  {Rikvold}(2015)}]{BuendiaPhyA}%
  \BibitemOpen
  \bibfield  {author} {\bibinfo {author} {\bibfnamefont {G.}~\bibnamefont
  {{Buend\'{\i}a}}}\ and\ \bibinfo {author} {\bibfnamefont {P.}~\bibnamefont
  {Rikvold}},\ }\href@noop {} {\bibfield  {journal} {\bibinfo  {journal} {Phys.
  A: Statistical Mechanics and its Applications}\ }\textbf {\bibinfo {volume}
  {424}},\ \bibinfo {pages} {217 } (\bibinfo {year} {2015})}\BibitemShut
  {NoStop}%
\bibitem [{\citenamefont {{Buend\'{\i}a}}\ and\ \citenamefont
  {Rikvold}(2013)}]{BuendiaPRE}%
  \BibitemOpen
  \bibfield  {author} {\bibinfo {author} {\bibfnamefont {G.~M.}\ \bibnamefont
  {{Buend\'{\i}a}}}\ and\ \bibinfo {author} {\bibfnamefont {P.~A.}\
  \bibnamefont {Rikvold}},\ }\href@noop {} {\bibfield  {journal} {\bibinfo
  {journal} {Phys. Rev. E}\ }\textbf {\bibinfo {volume} {88}},\ \bibinfo
  {pages} {012132} (\bibinfo {year} {2013})}\BibitemShut {NoStop}%
\bibitem [{\citenamefont {Bustos}\ \emph {et~al.}(2000)\citenamefont {Bustos},
  \citenamefont {U\~{n}ac},\ and\ \citenamefont {Zgrablich}}]{BustosPRE}%
  \BibitemOpen
  \bibfield  {author} {\bibinfo {author} {\bibfnamefont {V.}~\bibnamefont
  {Bustos}}, \bibinfo {author} {\bibfnamefont {R.~O.}\ \bibnamefont
  {U\~{n}ac}}, \ and\ \bibinfo {author} {\bibfnamefont {G.}~\bibnamefont
  {Zgrablich}},\ }\href@noop {} {\bibfield  {journal} {\bibinfo  {journal}
  {Phys. Rev. E}\ }\textbf {\bibinfo {volume} {62}},\ \bibinfo {pages} {8768}
  (\bibinfo {year} {2000})}\BibitemShut {NoStop}%
\bibitem [{\citenamefont {Mart\'{\i}n}\ \emph {et~al.}(2014)\citenamefont
  {Mart\'{\i}n}, \citenamefont {Bonachela},\ and\ \citenamefont
  {{Mu\~{n}oz}}}]{martin}%
  \BibitemOpen
  \bibfield  {author} {\bibinfo {author} {\bibfnamefont {P.~V.}\ \bibnamefont
  {Mart\'{\i}n}}, \bibinfo {author} {\bibfnamefont {J.~A.}\ \bibnamefont
  {Bonachela}}, \ and\ \bibinfo {author} {\bibfnamefont {M.~A.}\ \bibnamefont
  {{Mu\~{n}oz}}},\ }\href@noop {} {\bibfield  {journal} {\bibinfo  {journal}
  {Phys. Rev. E}\ }\textbf {\bibinfo {volume} {89}},\ \bibinfo {pages} {012145}
  (\bibinfo {year} {2014})}\BibitemShut {NoStop}%
\bibitem [{\citenamefont {Liu}\ \emph {et~al.}(2007)\citenamefont {Liu},
  \citenamefont {Guo},\ and\ \citenamefont {Evans}}]{Liu2007}%
  \BibitemOpen
  \bibfield  {author} {\bibinfo {author} {\bibfnamefont {D.-J.}\ \bibnamefont
  {Liu}}, \bibinfo {author} {\bibfnamefont {X.}~\bibnamefont {Guo}}, \ and\
  \bibinfo {author} {\bibfnamefont {J.~W.}\ \bibnamefont {Evans}},\ }\href
  {\doibase 10.1103/PhysRevLett.98.050601} {\bibfield  {journal} {\bibinfo
  {journal} {Phys. Rev. Lett.}\ }\textbf {\bibinfo {volume} {98}},\ \bibinfo
  {pages} {050601} (\bibinfo {year} {2007})}\BibitemShut {NoStop}%
\bibitem [{\citenamefont {Hilhorsta}(2008)}]{hilh08}%
  \BibitemOpen
  \bibfield  {author} {\bibinfo {author} {\bibfnamefont {H.}~\bibnamefont
  {Hilhorsta}},\ }\href {\doibase 10.1140/epjb/e2008-00003-7} {\bibfield
  {journal} {\bibinfo  {journal} {Eur. Phys. J. B}\ }\textbf {\bibinfo {volume}
  {64}},\ \bibinfo {pages} {437} (\bibinfo {year} {2008})}\BibitemShut
  {NoStop}%
\bibitem [{\citenamefont {Okabe}\ \emph {et~al.}(2000)\citenamefont {Okabe},
  \citenamefont {Boots}, \citenamefont {Sugihara},\ and\ \citenamefont
  {Chiu}}]{okabe}%
  \BibitemOpen
  \bibfield  {author} {\bibinfo {author} {\bibfnamefont {A.}~\bibnamefont
  {Okabe}}, \bibinfo {author} {\bibfnamefont {B.}~\bibnamefont {Boots}},
  \bibinfo {author} {\bibfnamefont {K.}~\bibnamefont {Sugihara}}, \ and\
  \bibinfo {author} {\bibfnamefont {S.~N.}\ \bibnamefont {Chiu}},\ }\href@noop
  {} {\emph {\bibinfo {title} {Spatial tessellations: concepts and applications
  of Voronoi diagrams}}}\ (\bibinfo  {publisher} {John Wiley and Sons Lts.},\
  \bibinfo {address} {Cichester},\ \bibinfo {year} {2000})\BibitemShut
  {NoStop}%
\bibitem [{\citenamefont {de~Oliveira}\ \emph {et~al.}(2008)\citenamefont
  {de~Oliveira}, \citenamefont {Alves}, \citenamefont {Ferreira},\ and\
  \citenamefont {Dickman}}]{oliveira2}%
  \BibitemOpen
  \bibfield  {author} {\bibinfo {author} {\bibfnamefont {M.~M.}\ \bibnamefont
  {de~Oliveira}}, \bibinfo {author} {\bibfnamefont {S.~G.}\ \bibnamefont
  {Alves}}, \bibinfo {author} {\bibfnamefont {S.~C.}\ \bibnamefont {Ferreira}},
  \ and\ \bibinfo {author} {\bibfnamefont {R.}~\bibnamefont {Dickman}},\ }\href
  {\doibase 10.1103/PhysRevE.78.031133} {\bibfield  {journal} {\bibinfo
  {journal} {Phys. Rev. E}\ }\textbf {\bibinfo {volume} {78}},\ \bibinfo
  {pages} {031133} (\bibinfo {year} {2008})}\BibitemShut {NoStop}%
\bibitem [{\citenamefont {Luck}(1993)}]{luck93}%
  \BibitemOpen
  \bibfield  {author} {\bibinfo {author} {\bibfnamefont {J.~M.}\ \bibnamefont
  {Luck}},\ }\href@noop {} {\bibfield  {journal} {\bibinfo  {journal}
  {Europhys. Lett.}\ }\textbf {\bibinfo {volume} {24}},\ \bibinfo {pages} {359}
  (\bibinfo {year} {1993})}\BibitemShut {NoStop}%
\bibitem [{\citenamefont {Janke}\ and\ \citenamefont {Weigel}(2004)}]{janke}%
  \BibitemOpen
  \bibfield  {author} {\bibinfo {author} {\bibfnamefont {W.}~\bibnamefont
  {Janke}}\ and\ \bibinfo {author} {\bibfnamefont {M.}~\bibnamefont {Weigel}},\
  }\href {\doibase 10.1103/PhysRevB.69.144208} {\bibfield  {journal} {\bibinfo
  {journal} {Phys. Rev. B}\ }\textbf {\bibinfo {volume} {69}},\ \bibinfo
  {pages} {144208} (\bibinfo {year} {2004})}\BibitemShut {NoStop}%
\bibitem [{\citenamefont {Barghathi}\ and\ \citenamefont
  {Vojta}(2014)}]{vojta14c}%
  \BibitemOpen
  \bibfield  {author} {\bibinfo {author} {\bibfnamefont {H.}~\bibnamefont
  {Barghathi}}\ and\ \bibinfo {author} {\bibfnamefont {T.}~\bibnamefont
  {Vojta}},\ }\href@noop {} {\bibfield  {journal} {\bibinfo  {journal} {Phys.
  Rev. Lett.}\ }\textbf {\bibinfo {volume} {113}},\ \bibinfo {pages} {120602}
  (\bibinfo {year} {2014})}\BibitemShut {NoStop}%
\bibitem [{\citenamefont {Turcotte}(1999)}]{Turcotte1999}%
  \BibitemOpen
  \bibfield  {author} {\bibinfo {author} {\bibfnamefont {D.~L.}\ \bibnamefont
  {Turcotte}},\ }\href {\doibase 10.1088/0034-4885/62/10/201} {\bibfield
  {journal} {\bibinfo  {journal} {Rep. Prog. Phys.}\ }\textbf {\bibinfo
  {volume} {62}},\ \bibinfo {pages} {1377} (\bibinfo {year}
  {1999})}\BibitemShut {NoStop}%
\bibitem [{\citenamefont {Friedberg}\ and\ \citenamefont
  {Ren}(1984)}]{Friedberg1984}%
  \BibitemOpen
  \bibfield  {author} {\bibinfo {author} {\bibfnamefont {R.}~\bibnamefont
  {Friedberg}}\ and\ \bibinfo {author} {\bibfnamefont {H.-C.}\ \bibnamefont
  {Ren}},\ }\href {\doibase 10.1016/0550-3213(84)90501-7} {\bibfield  {journal}
  {\bibinfo  {journal} {Nucl. Phys. B}\ }\textbf {\bibinfo {volume} {235}},\
  \bibinfo {pages} {310} (\bibinfo {year} {1984})}\BibitemShut {NoStop}%
\bibitem [{\citenamefont {Harris}(1974{\natexlab{b}})}]{harris-CP}%
  \BibitemOpen
  \bibfield  {author} {\bibinfo {author} {\bibfnamefont {T.~E.}\ \bibnamefont
  {Harris}},\ }\href@noop {} {\bibfield  {journal} {\bibinfo  {journal} {Ann.
  Probab.}\ }\textbf {\bibinfo {volume} {2}},\ \bibinfo {pages} {969} (\bibinfo
  {year} {1974}{\natexlab{b}})}\BibitemShut {NoStop}%
\bibitem [{\citenamefont {Dickman}\ and\ \citenamefont {Burschka}(1987)}]{a1}%
  \BibitemOpen
  \bibfield  {author} {\bibinfo {author} {\bibfnamefont {R.}~\bibnamefont
  {Dickman}}\ and\ \bibinfo {author} {\bibfnamefont {M.}~\bibnamefont
  {Burschka}},\ }\href@noop {} {\bibfield  {journal} {\bibinfo  {journal}
  {Phys. Lett. A}\ }\textbf {\bibinfo {volume} {127}},\ \bibinfo {pages} {132}
  (\bibinfo {year} {1987})}\BibitemShut {NoStop}%
\bibitem [{\citenamefont {Dickman}\ and\ \citenamefont {Kamphorst Leal~da
  Silva}(1998)}]{dic-jaf}%
  \BibitemOpen
  \bibfield  {author} {\bibinfo {author} {\bibfnamefont {R.}~\bibnamefont
  {Dickman}}\ and\ \bibinfo {author} {\bibfnamefont {J.}~\bibnamefont
  {Kamphorst Leal~da Silva}},\ }\href {\doibase 10.1103/PhysRevE.58.4266}
  {\bibfield  {journal} {\bibinfo  {journal} {Phys. Rev. E}\ }\textbf {\bibinfo
  {volume} {58}},\ \bibinfo {pages} {4266} (\bibinfo {year}
  {1998})}\BibitemShut {NoStop}%
\bibitem [{\citenamefont {Vojta}\ and\ \citenamefont {Hoyos}(2014)}]{vojta14}%
  \BibitemOpen
  \bibfield  {author} {\bibinfo {author} {\bibfnamefont {T.}~\bibnamefont
  {Vojta}}\ and\ \bibinfo {author} {\bibfnamefont {J.~A.}\ \bibnamefont
  {Hoyos}},\ }\href {\doibase 10.1103/PhysRevLett.112.075702} {\bibfield
  {journal} {\bibinfo  {journal} {Phys. Rev. Lett.}\ }\textbf {\bibinfo
  {volume} {112}},\ \bibinfo {pages} {075702} (\bibinfo {year}
  {2014})}\BibitemShut {NoStop}%
\bibitem [{\citenamefont {Jensen}\ and\ \citenamefont {Dickman}(1994)}]{a2}%
  \BibitemOpen
  \bibfield  {author} {\bibinfo {author} {\bibfnamefont {I.}~\bibnamefont
  {Jensen}}\ and\ \bibinfo {author} {\bibfnamefont {R.}~\bibnamefont
  {Dickman}},\ }\href@noop {} {\bibfield  {journal} {\bibinfo  {journal} {Phys.
  A}\ }\textbf {\bibinfo {volume} {203}},\ \bibinfo {pages} {175} (\bibinfo
  {year} {1994})}\BibitemShut {NoStop}%
\bibitem [{\citenamefont {Dickman}\ \emph {et~al.}(2001)\citenamefont
  {Dickman}, \citenamefont {Alava}, \citenamefont {{A. Mu\~{n}oz}},
  \citenamefont {Peltola}, \citenamefont {Vespignani},\ and\ \citenamefont
  {Zapperi}}]{Dickman2001}%
  \BibitemOpen
  \bibfield  {author} {\bibinfo {author} {\bibfnamefont {R.}~\bibnamefont
  {Dickman}}, \bibinfo {author} {\bibfnamefont {M.}~\bibnamefont {Alava}},
  \bibinfo {author} {\bibfnamefont {M.}~\bibnamefont {{A. Mu\~{n}oz}}},
  \bibinfo {author} {\bibfnamefont {J.}~\bibnamefont {Peltola}}, \bibinfo
  {author} {\bibfnamefont {A.}~\bibnamefont {Vespignani}}, \ and\ \bibinfo
  {author} {\bibfnamefont {S.}~\bibnamefont {Zapperi}},\ }\href {\doibase
  10.1103/PhysRevE.64.056104} {\bibfield  {journal} {\bibinfo  {journal} {Phys.
  Rev. E}\ }\textbf {\bibinfo {volume} {64}},\ \bibinfo {pages} {056104}
  (\bibinfo {year} {2001})},\ \Eprint {http://arxiv.org/abs/0101381} {0101381}
  \BibitemShut {NoStop}%
\bibitem [{\citenamefont {de~Oliveira}\ and\ \citenamefont
  {Dickman}(2005)}]{qssimPRE}%
  \BibitemOpen
  \bibfield  {author} {\bibinfo {author} {\bibfnamefont {M.~M.}\ \bibnamefont
  {de~Oliveira}}\ and\ \bibinfo {author} {\bibfnamefont {R.}~\bibnamefont
  {Dickman}},\ }\href {\doibase 10.1103/PhysRevE.71.016129} {\bibfield
  {journal} {\bibinfo  {journal} {Phys. Rev. E}\ }\textbf {\bibinfo {volume}
  {71}},\ \bibinfo {pages} {016129} (\bibinfo {year} {2005})}\BibitemShut
  {NoStop}%
\bibitem [{\citenamefont {Dickman}\ and\ \citenamefont
  {de~Oliveira}(2005)}]{qssimPhysA}%
  \BibitemOpen
  \bibfield  {author} {\bibinfo {author} {\bibfnamefont {R.}~\bibnamefont
  {Dickman}}\ and\ \bibinfo {author} {\bibfnamefont {M.~M.}\ \bibnamefont
  {de~Oliveira}},\ }\href@noop {} {\bibfield  {journal} {\bibinfo  {journal}
  {Phys. A}\ }\textbf {\bibinfo {volume} {357}},\ \bibinfo {pages} {134 }
  (\bibinfo {year} {2005})}\BibitemShut {NoStop}%
\bibitem [{\citenamefont {Binder}(1987)}]{binder}%
  \BibitemOpen
  \bibfield  {author} {\bibinfo {author} {\bibfnamefont {K.}~\bibnamefont
  {Binder}},\ }\href {http://stacks.iop.org/0034-4885/50/i=7/a=001} {\bibfield
  {journal} {\bibinfo  {journal} {Reports on Progress in Physics}\ }\textbf
  {\bibinfo {volume} {50}},\ \bibinfo {pages} {783} (\bibinfo {year}
  {1987})}\BibitemShut {NoStop}%
\bibitem [{\citenamefont {{Ali Saif}}\ and\ \citenamefont
  {Gade}(2009)}]{AliSaif2009}%
  \BibitemOpen
  \bibfield  {author} {\bibinfo {author} {\bibfnamefont {M.}~\bibnamefont {{Ali
  Saif}}}\ and\ \bibinfo {author} {\bibfnamefont {P.~M.}\ \bibnamefont
  {Gade}},\ }\href {\doibase 10.1088/1742-5468/2009/07/P07023} {\bibfield
  {journal} {\bibinfo  {journal} {J. Stat. Mech.: Theor. Exp.}\ ,\ \bibinfo
  {pages} {P07023}} (\bibinfo {year} {2009})}\BibitemShut {NoStop}%
\bibitem [{\citenamefont {Sinha}\ and\ \citenamefont
  {Mukherjee}(2012)}]{Sinha2012}%
  \BibitemOpen
  \bibfield  {author} {\bibinfo {author} {\bibfnamefont {I.}~\bibnamefont
  {Sinha}}\ and\ \bibinfo {author} {\bibfnamefont {A.~K.}\ \bibnamefont
  {Mukherjee}},\ }\href {\doibase 10.1007/s10955-011-0414-5} {\bibfield
  {journal} {\bibinfo  {journal} {J. Stat. Phys.}\ }\textbf {\bibinfo {volume}
  {146}},\ \bibinfo {pages} {669} (\bibinfo {year} {2012})}\BibitemShut
  {NoStop}%
\bibitem [{\citenamefont {de~Oliveira}\ \emph {et~al.}(2015)\citenamefont
  {de~Oliveira}, \citenamefont {da~Luz},\ and\ \citenamefont
  {Fiore}}]{DeOliveira2015}%
  \BibitemOpen
  \bibfield  {author} {\bibinfo {author} {\bibfnamefont {M.~M.}\ \bibnamefont
  {de~Oliveira}}, \bibinfo {author} {\bibfnamefont {M.~G.~E.}\ \bibnamefont
  {da~Luz}}, \ and\ \bibinfo {author} {\bibfnamefont {C.~E.}\ \bibnamefont
  {Fiore}},\ }\href@noop {} {\bibfield  {journal} {\bibinfo  {journal} {Phys.
  Rev. E}\ }\textbf {\bibinfo {volume} {92}},\ \bibinfo {pages} {062126}
  (\bibinfo {year} {2015})}\BibitemShut {NoStop}%
\bibitem [{\citenamefont {L\"{u}beck}(2002)}]{Lubeck2002}%
  \BibitemOpen
  \bibfield  {author} {\bibinfo {author} {\bibfnamefont {S.}~\bibnamefont
  {L\"{u}beck}},\ }\href {\doibase 10.1103/PhysRevE.66.046114} {\bibfield
  {journal} {\bibinfo  {journal} {Phys. Rev. E}\ }\textbf {\bibinfo {volume}
  {66}},\ \bibinfo {pages} {046114} (\bibinfo {year} {2002})}\BibitemShut
  {NoStop}%
\bibitem [{\citenamefont {da~Cunha}\ \emph {et~al.}(2014)\citenamefont
  {da~Cunha}, \citenamefont {da~Silva}, \citenamefont {Viswanathan},\ and\
  \citenamefont {Dickman}}]{DaCunha2014}%
  \BibitemOpen
  \bibfield  {author} {\bibinfo {author} {\bibfnamefont {S.~D.}\ \bibnamefont
  {da~Cunha}}, \bibinfo {author} {\bibfnamefont {L.~R.}\ \bibnamefont
  {da~Silva}}, \bibinfo {author} {\bibfnamefont {G.~M.}\ \bibnamefont
  {Viswanathan}}, \ and\ \bibinfo {author} {\bibfnamefont {R.}~\bibnamefont
  {Dickman}},\ }\href {\doibase 10.1088/1742-5468/2014/08/P08003} {\bibfield
  {journal} {\bibinfo  {journal} {J. Stat. Mech.: Theor. Exp.}\ ,\ \bibinfo
  {pages} {P08003}} (\bibinfo {year} {2014})}\BibitemShut {NoStop}%
\bibitem [{\citenamefont {Bonachela}\ and\ \citenamefont
  {Mu\~{n}oz}(2007)}]{Bonachela2007}%
  \BibitemOpen
  \bibfield  {author} {\bibinfo {author} {\bibfnamefont {J.~A.}\ \bibnamefont
  {Bonachela}}\ and\ \bibinfo {author} {\bibfnamefont {M.~A.}\ \bibnamefont
  {Mu\~{n}oz}},\ }\href {\doibase 10.1016/j.physa.2007.04.110} {\bibfield
  {journal} {\bibinfo  {journal} {Phys. A}\ }\textbf {\bibinfo {volume}
  {384}},\ \bibinfo {pages} {89} (\bibinfo {year} {2007})}\BibitemShut
  {NoStop}%
\bibitem [{\citenamefont {Dickman}(2006)}]{Dickman2006}%
  \BibitemOpen
  \bibfield  {author} {\bibinfo {author} {\bibfnamefont {R.}~\bibnamefont
  {Dickman}},\ }\href {\doibase 10.1103/PhysRevE.73.036131} {\bibfield
  {journal} {\bibinfo  {journal} {Phys. Rev. E}\ }\textbf {\bibinfo {volume}
  {73}},\ \bibinfo {pages} {036131} (\bibinfo {year} {2006})}\BibitemShut
  {NoStop}%
\end{thebibliography}

%

\end{document}